\documentclass[conference]{IEEEtran}[a4]
\IEEEoverridecommandlockouts
\usepackage{graphicx}
\usepackage{subfigure}
\usepackage{url}
\RequirePackage[OT1]{fontenc}

\usepackage{multirow}
\usepackage{hyperref}

\newcommand{\cc}{Cool-Chic~}

\begin{document}

\title{Upsampling Improvement for Overfitted Neural Coding }

\author{\IEEEauthorblockN{Pierrick Philippe, Th\'eo Ladune, Gordon Clare, F\'elix Henry, Th\'eophile Blard and Thomas Leguay}
\IEEEauthorblockA{Orange Innovation, France \\
\texttt{firstname.lastname@orange.com}}
}

\maketitle

\begin{abstract}

Neural image compression, based on auto-encoders and overfitted representations,  relies on a latent representation of the coded signal. This representation needs to be compact and uses low resolution feature maps. In the decoding process, those latents are upsampled and filtered using stacks of convolution filters and non linear elements to recover the decoded image.

Therefore, the upsampling process is crucial in the design of a neural coding scheme and is of particular importance for overfitted codecs where the network parameters, including the upsampling filters,  are part of the representation.

This paper addresses the improvement of the upsampling process in order to reduce its complexity and limit the number of parameters. A new upsampling structure is presented whose improvements are illustrated within the \cc overfitted image coding framework. The proposed approach offers a rate reduction of 4.7\%. The source code is available~\cite{ccopen}.

\end{abstract}


\section{Introduction \& Related work}

Image compression  mainly relies on transform coding to provide a compact representation of the signal.
Indeed, the transform domain is the place where the actual compression takes place, through quantisation and entropy (lossless) coding.
In conventional technologies, such as JPEG and HEIF~(HEVC)~\cite{hevc}, the transform or frequency domain is obtained through trigonometric transforms such as the DCT or the DST.
For JPEG 2000~\cite{10.5555/559856}, the image signal is compressed through a wavelet transform, and is then organised in a pyramidal and frequency fashion.

For neural coding, such as JPEG-AI~\cite{10123093},~\cite{balle}, the transform domain, often called latent representation, is learned through a rate distortion optimisation process and automatically provides a compact signal representation.

In both cases, the latent/transform domain is in the downsampled domain, i.e. the compressed signal representation involves numerous channels with resolutions lower than the initial image resolution. As an example~\cite{balle}, propose a latent representation of typically 192 latents with $W/16 \times H/16$  values, where $W$ and $H$ stand for the width and height of the coded image.

At the decoding side, the upsampling process, that aims at recovering the image resolution, merely consist in successive transpose convolutions with a set of $\times 2$ upsampling cells, that is a stride=2. This upsampling process is learned once for all, and usually is not limited by complexity. Consequently, most of the time the upsampling kernels, although of limited size e.g. $5\times 5$, involve 2D convolutions with no additional constraints.

Recently, overfitted neural coding has been introduced~\cite{dupont2021coincompressionimplicitneural},\cite{coolchic} where the neural network is adapted to the coded image. For COIN it solely carries the image, and a coordinate based decoding through a synthesis stage is invoked for decoding. In the case of \cc, a latent is also conveyed along with the neural network, and this latent representation is decoded using the synthesis. As the latent representation is based on downsampled channels, the synthesis involves upsampling.

\cc  upsampling and synthesis parameters are part of the image representation and need to be compact. Therefore,  the number of parameters for the upsampler has to be limited as much as possible. Also, considering the complexity of the decoding process, the upsampling process complexity must remain low to reduce the overall number of multiplication accumulation operations (MAC). This is the reason why in~\cite{blard}, the upsampling is parameterised and the filter taps reduced to address lower complexity decoders.

As such, for lower complexity levels, \cc relies on a learned linear upsampling process with a reduced $4\times 4$ kernel, also limiting the number of filter taps to convey.

Although \cc performance is comparable to the VVC standard~\cite{9503377} for RGB pictures, the upsampling presents some limitations that are considered in this paper.

First, as the filters are non-separable, they exhibit an implementation complexity and a number of taps which is quadratic with the filter order. In the context of low complexity overfitted codecs a separable structure is advisable as it is less complex and the number of taps can be reduced to allow additional flexibility:  Indeed, we show that the adoption of separable filters allows an increase of the number of filter taps and an improvement of the overfitting process through better kernel adaptations.

It is also shown that, through the addition of an additional low complexity filter, the upsampling and \cc performance is increased.

The paper is organised as follows:
\begin{itemize}
\item  First, an introduction to the \cc codec and the upsampling process is made.
\item a new filtering structure is introduced to overcome the limitation of the legacy implementation. A separable, symmetric and novel upsampling structure is presented.
\item the benefits of the proposed architecture and the additional levels of flexibility are highlighted through experiments where noticeable coding improvement is demonstrated.
\end{itemize}

\section{Low complexity overfitted image coding}

Recently the field of visual representation has seen a new paradigm: Implicit Neural Representation aims at describing a visual scene with an overfitted neural network. In that context~\cite{dupont2021coincompressionimplicitneural} introduces a neural decoder based on coordinates in which the neural network maps pixel locations to RGB values for a single image. During a learning process, the neural network is overfitted to the image, as such it carries the image representation.

This work was further extended in~\cite{coolchic} and ~\cite{c3} where signal latents are introduced along with the overfitted network: the latent representation is learned in conjunction with a neural decoder to provide the image representation. The coded image is then composed of the coded entropy coded latents and neural network weights.

\subsection{Brief description of \cc}

The key objective of \cc is to provide a low complexity decoder whose weights are trained in an end-to-end fashion. The quantised latents are learned alongside the decoder weights in a rate distortion manner, i.e. the quantity $J(\lambda)=D+ \lambda \cdot R$ is minimised. Any derivable distortion $D$ can be used in that context as shown in~\cite{10533758}, $R$ is the number of bits per pixel and $\lambda$ balances the two quantities.

\cc surpasses VVC image coding efficiency in the Mean Square Error domain as reported in \href{https://orange-opensource.github.io/Cool-Chic/}{its open source implementation}. Several decoder architectures are suggested for a decoding complexity in a range of 300 to 2300 MAC per decoded pixel.

A \cc decoder is made of four elements:
\begin{enumerate}
\item a series of L=7 latents, organised in a pyramidal fashion: their resolution ranges from 1 to 1/64 of the image
\item an entropy decoder, based on an autoregressive network, which provides a statistical description of the latents. These statistics drive a binary arithmetic decoder
\item a linear upsampling block, built with a series of $\times 2$ upsamplers to convert the pyramidal latent to a dense representation for the L latents, at the image resolution
\item a synthesis neural network which takes the resulting dense latents and outputs the RGB image color channels.
\end{enumerate}

As a function of the desired complexity, the autoregressive and the synthesis networks complexity is adjusted adding more neurons and hidden layers.

\subsection{Upsampling details and limitations}

\begin{figure}
\begin{center}
    \includegraphics[width=\columnwidth]{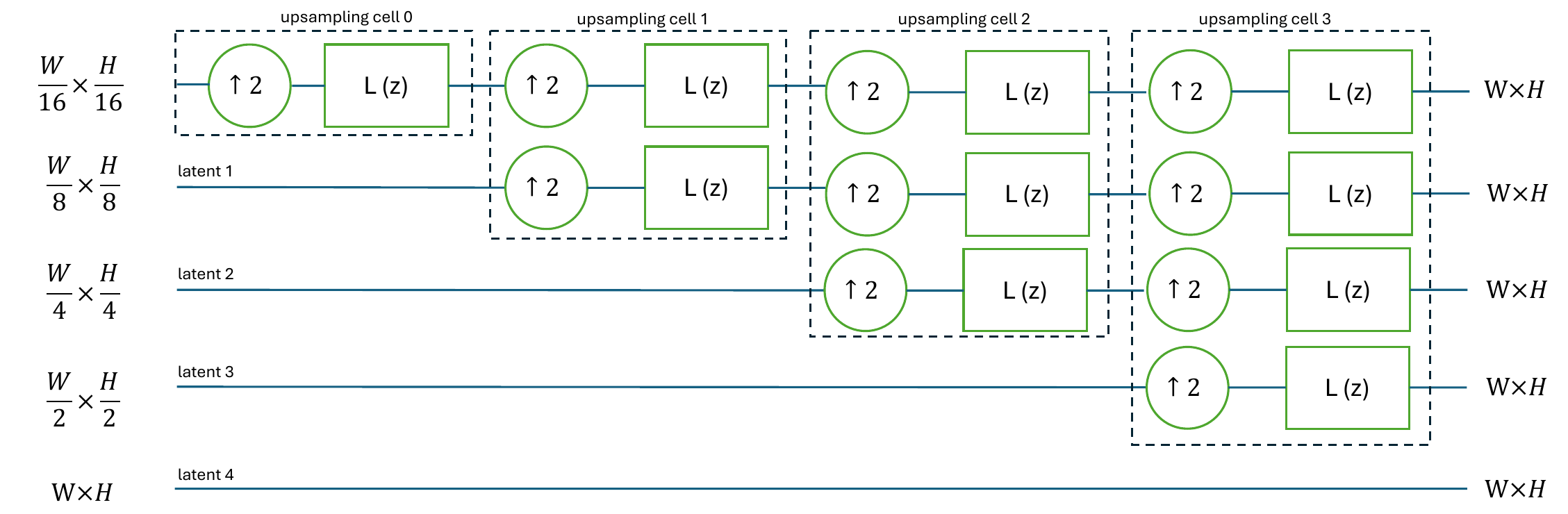}
	\caption{\cc upsampling process, L=5 latents are presented}
	\label{fig:ups_cc}
\end{center}
\end{figure}

The \cc upsampling process is detailed figure~\ref{fig:ups_cc}. The pyramidal latents are progressively upsampled by groups of upsamplers to recover a dense latent representation, i.e. a tensor of $L\times W\times H$ values. The kernel size for the upsampler $L(z)$ is set to $K\times K$ with $K=4$ or $K=8$ depending on the \cc complexity.

The upsampler complexity can be calculated accounting for the number of downsamplers and the latent resolutions, calculating:

\begin{equation}
\mathcal{O}(K) = K^2 \cdot \sum_{n=1}^{n=6} \frac{n}{2^n}
\end{equation}

This implies respectively 30 and 121 MAC per decoded pixel for $K=4$ and $K=8$.

\begin{figure*}
\centering
\begin{subfigure}[]	{\includegraphics[width=0.18\textwidth]{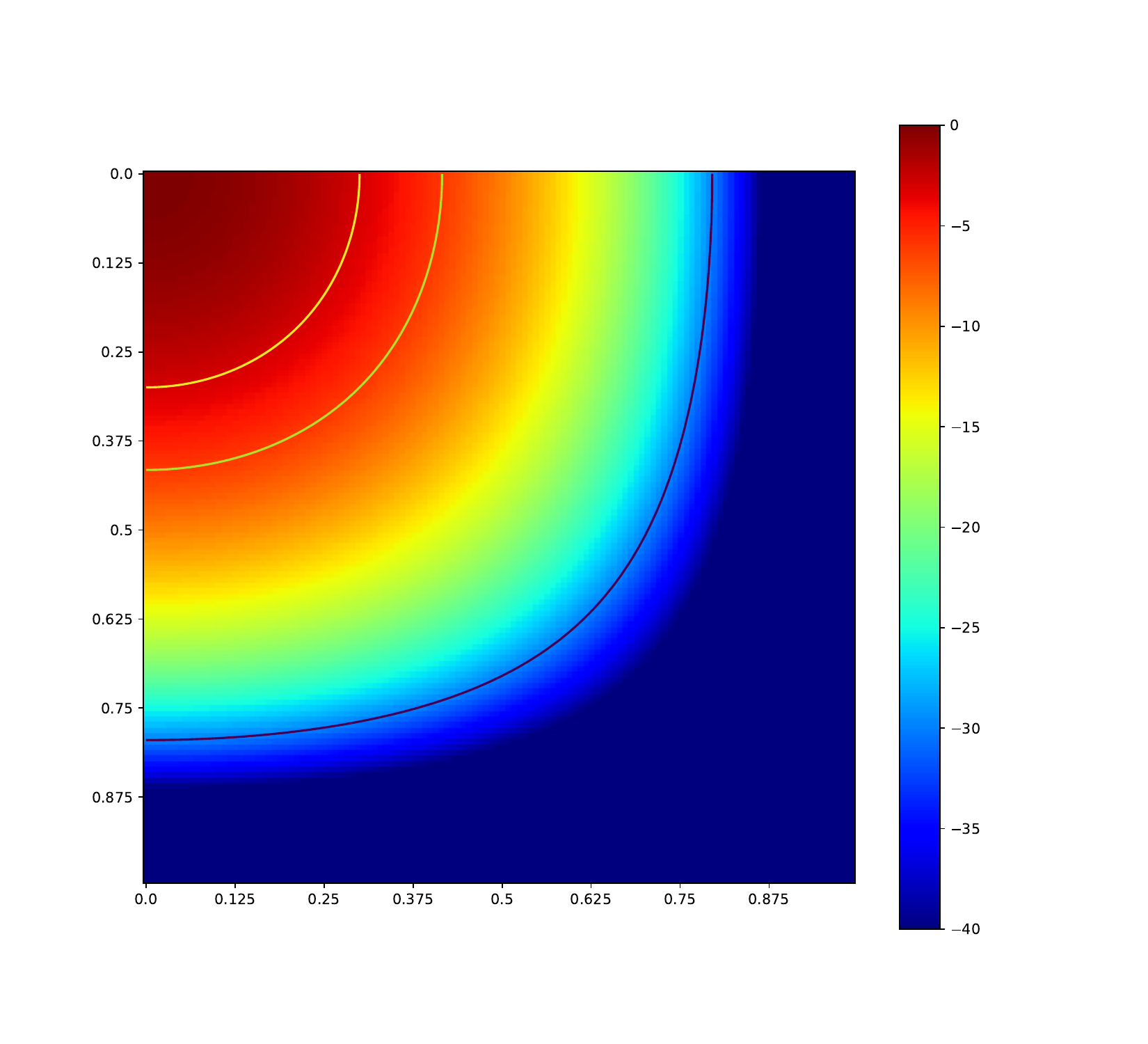}}
\end{subfigure}
\begin{subfigure}[]	{\includegraphics[width=0.18\textwidth]{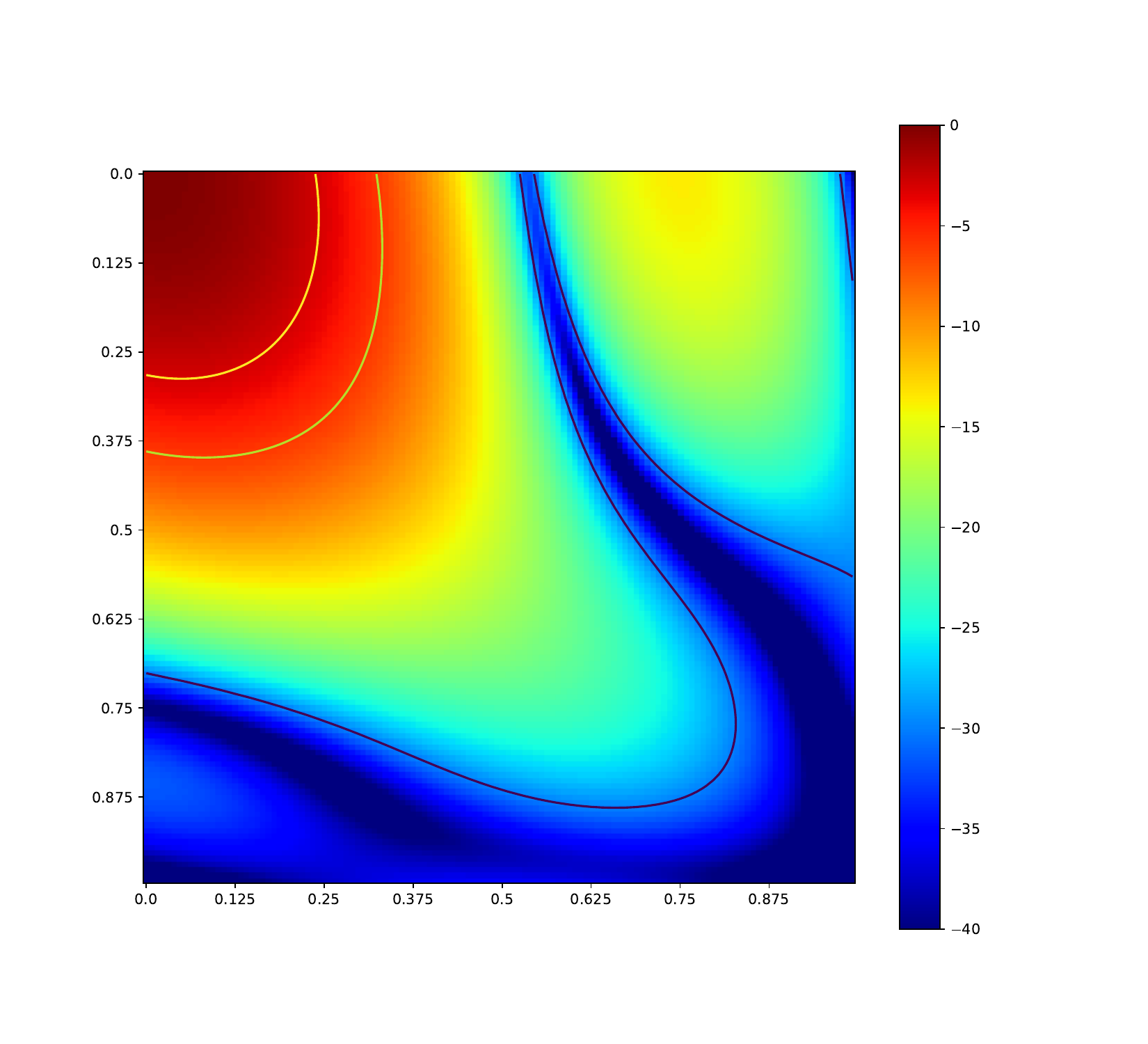}}
\end{subfigure}
\begin{subfigure}[]	{\includegraphics[width=0.18\textwidth]{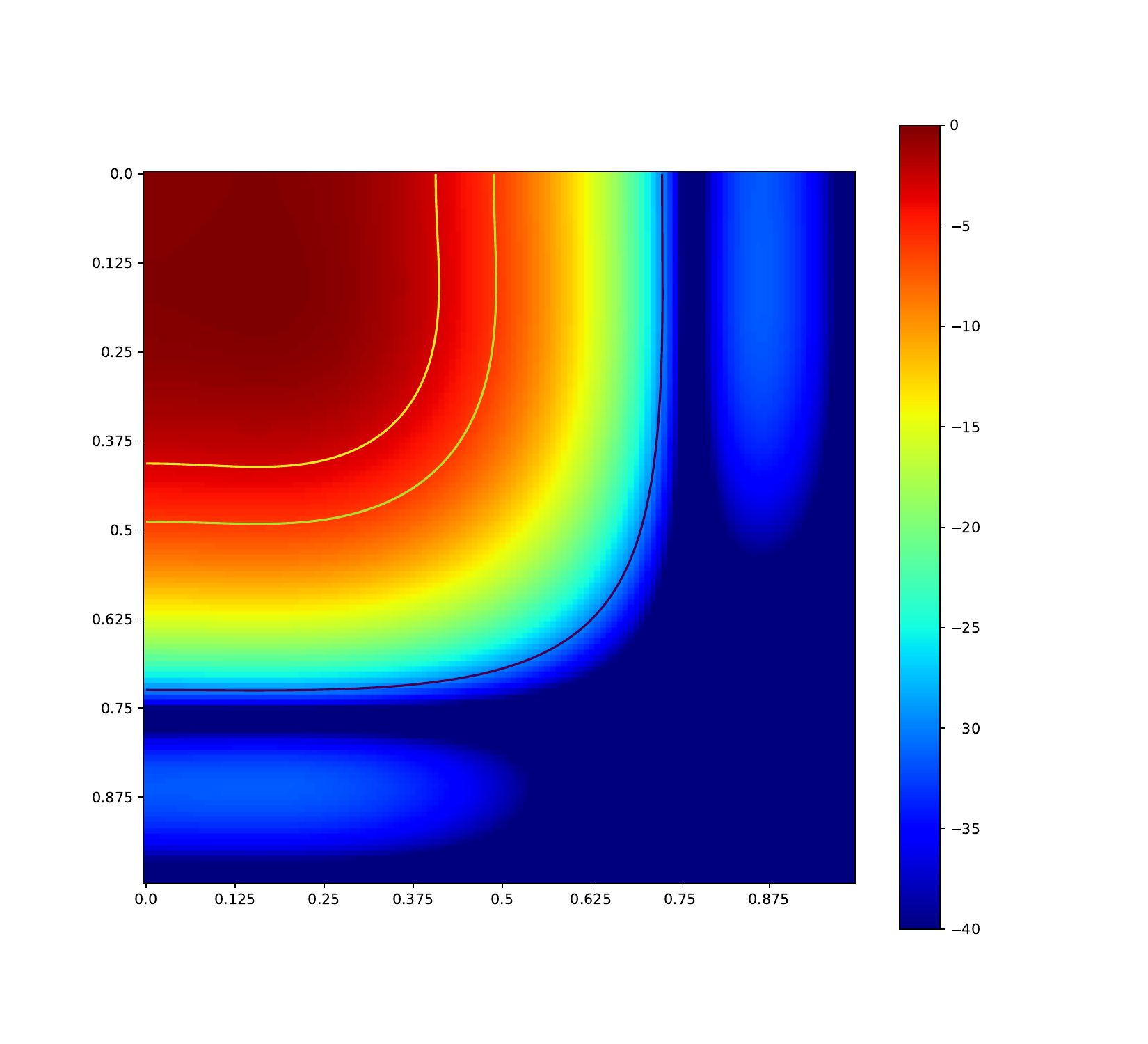}}
\end{subfigure}
\begin{subfigure}[]	{\includegraphics[width=0.18\textwidth]{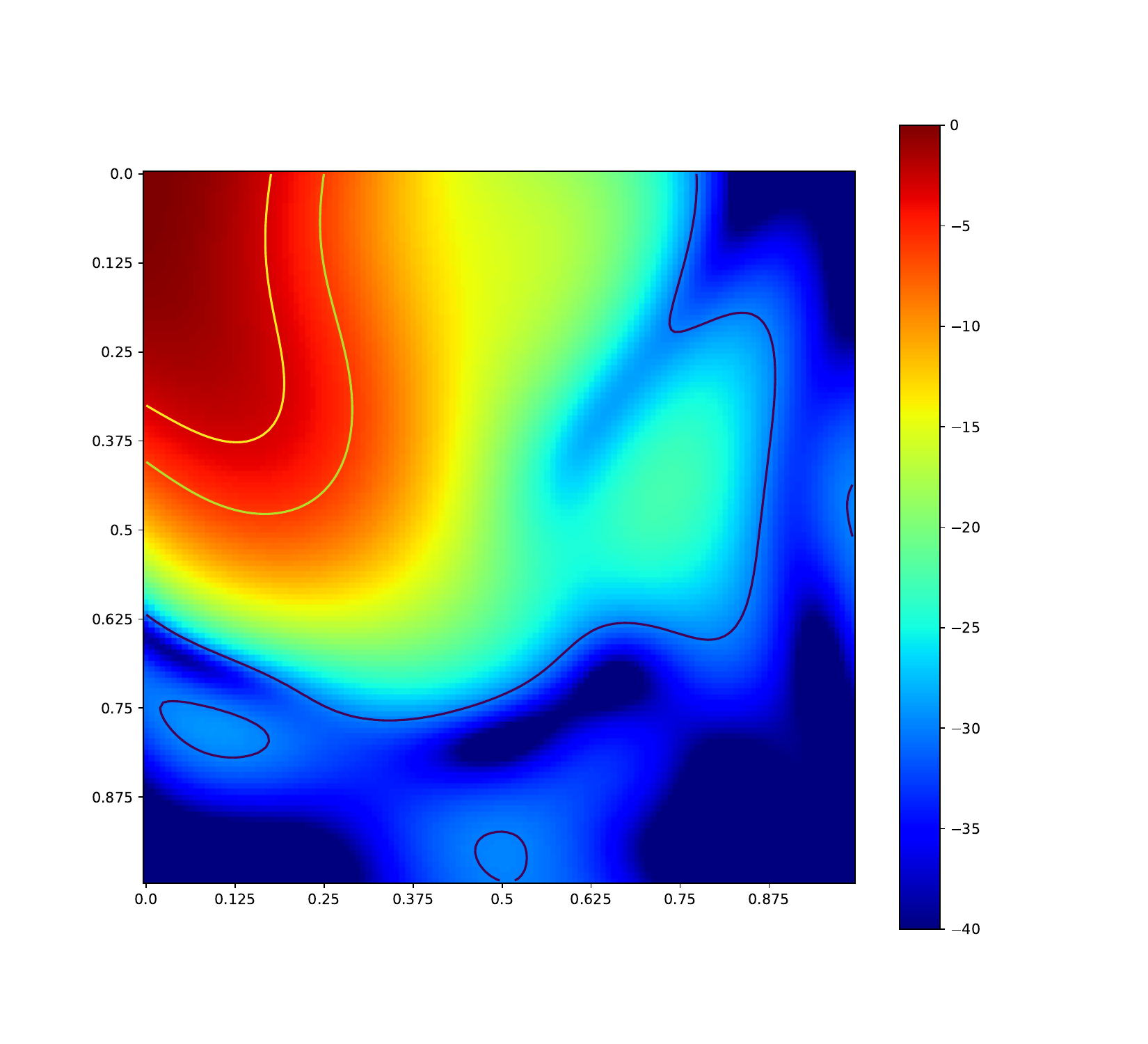}}
\end{subfigure}
\begin{subfigure}[]	{\includegraphics[width=0.18\textwidth]{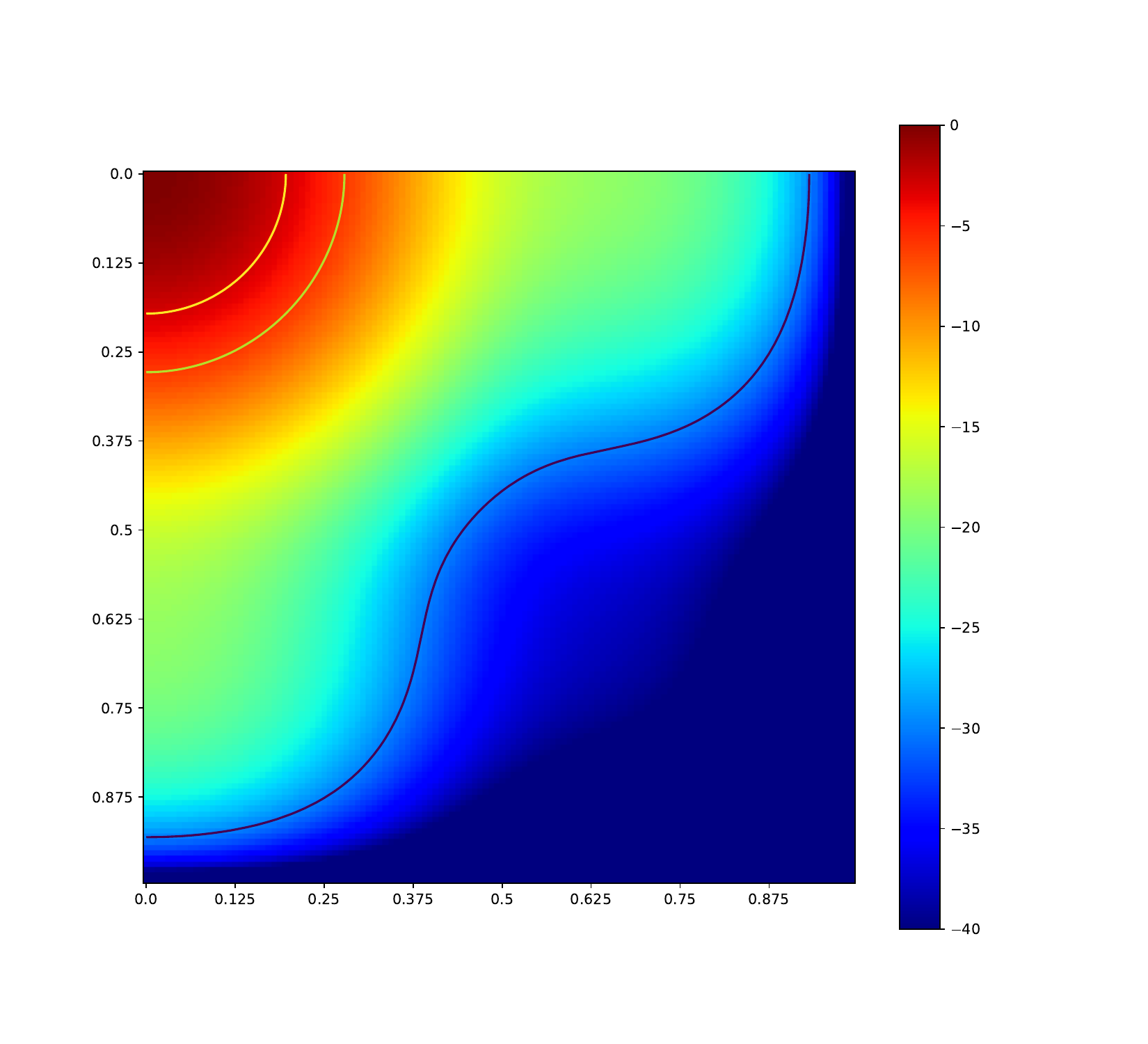}}
\end{subfigure}
\caption{Case $K=4$: Magnitude of the frequency response of a bilinear filter (a) and the optimised filter (b). Case $K=8$: For a bicubic filter (c) and the optimised filter (d). When turned into separable and symmetrical the learned filter converges to (e). The frequency is normalised and the attenuation in dB is reported on the colorbar. The attenuation at -3dB and -6dB is reported as a yellow line.}
\label{fig:cc_freqresponse}
\end{figure*}

In the case of $K=4$, the upsampler is initialised with a bilinear kernel, for $K=8$ a bicubic kernel is used.
In the process of training the \cc decoder and corresponding latents, the upsampler taps are adjusted for each sequence during the learning process. Figure~\ref{fig:cc_freqresponse} presents the result of the optimisation process for the sequence sergey-zolkin-21232 of the CLIC-20 dataset for a 0.02 bpp bit budget. The frequency response modules $| L \left( f_1,f_2 \right) |$ are shown for respectively $K=4$ and $K=8$ at initialisation (bilinear / bicubic) and after optimisation.

The figure reveals that  the frequency response is strengthened, as the cut-off frequency (materialised by the yellow line at -3 dB and -6dB) shifts toward 0.25 after optimisation. Also, the frequency response appears to be slightly asymmetrical ($|L\left(f_1,f_2\right)| \neq \|L\left(f_2,f_1\right)|$): the learned filters are not separable.

\section{Proposed upsampling improvements}

\subsection{Separability and symmetry}

Although non-separable kernels are historically the default choice for learning based upsamplers, they reveal some drawbacks:

\begin{enumerate}
\item in the context of the serial upsampling process, their algorithmic complexity is quadratic with the kernel size $K$
\item they require $K^2$ parameters, this is a problem in the overfitted approach as all the parameters, including filter taps, are part of the image representation
\item also, as the number of parameters is constrained by the rate, the expressiveness of the network is de facto reduced as the number of filters cannot be made different e.g. for each upsampler level.
\end{enumerate}

Those reasons invite the use of separable filters in the upsampling process. Considering point 1) above, the complexity of the upsampling process reduces significantly as it rewrites:

\begin{equation}
\mathcal{O}(K) = 3K \cdot \sum_{n=1}^{n=6} \frac{n}{2^n}
\end{equation}

This reduces the number of MAC to respectively 23 and 45 MAC for $K=4$ and $K=8$. Although it doesn't affect the number of operations of the upsampling process, the separable filters are made identical for the vertical and horizontal application, and they are also chosen symmetrical. This has a significant advantage with respect to the number of parameters: for $K=8$, they are decreased from 64 to 4. This means that the number of independent filter cells can be increased leading to more adaptivity.

\subsection{Additional branch}

In figure~\ref{fig:ups_cc}, one sees that after each upsampling cell, the latent with a compatible resolution is brought to the subsequent upsampler. For example, latent 1 is input to the upsampling cell 1, etc. Comparing the reconstruction process of \cc with the one of a wavelet synthesis, it is noticed that the latent with the higher frequency content is not filtered in the current \cc design.

\begin{figure}
\centering
\begin{subfigure}[]	{\includegraphics[width=0.22\textwidth]{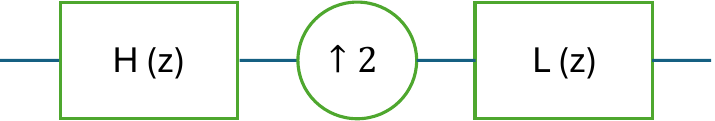}}\end{subfigure}
\begin{subfigure}[]	{\includegraphics[width=0.22\textwidth]{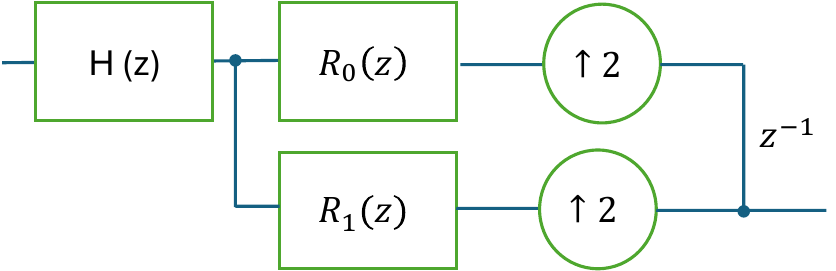}}\end{subfigure}
\caption{The two filtering structures are mathematically equivalent: the combination of filters $H$ and upsampling $L$ can be seen as a lengthening of the interpolation filter.}
\label{fig:polyphase}
\end{figure}

An additional set of filters  $H(z)$ is consequently added to ease the reconstruction process. This also can be seen as an additional means to specialise the filtering cell to the input latent. Indeed, following the polyphase decomposition~\ref{fig:polyphase}, the filter $H(z)$ can be seen as an extension of the polyphase decomposition of $L(z) = R_1(z^2)+ z^{-1}\cdot R_0(z^2)$. As $H(z)$ is also applied in the downsampled domain, this provides an additional overfitting possibility, maintaining reduced complexity as it is applied prior to the upsampler.

In order to reduce the number of parameters, $H(z)$ is also made separable and symmetric. At initialisation the filter is set to a 5-tap Dirac impulse.

\begin{figure}
\begin{center}
    \includegraphics[width=\columnwidth]{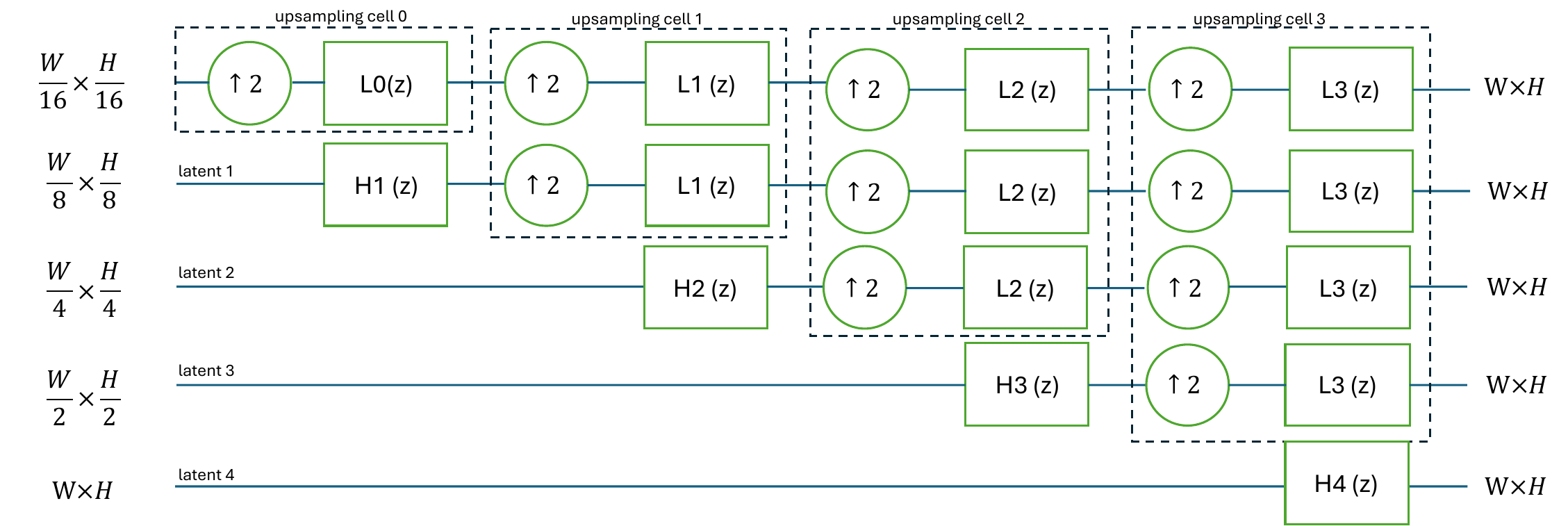}
	\caption{\cc upsampling process, L=5 latents are presented}
	\label{fig:ups_prop}
\end{center}
\end{figure}

The updated upsampling process is presented figure~\ref{fig:ups_prop}. As the filtering kernels are symmetric and separable and are more compact, their number is extended and $L$ interpolators $L_n(z)$ and $H_n(z)$ are provided.

In the next section, the benefits of the new structure is investigated through the influence of kernel sizes and number of interpolators.

\section{Results and experiments}

In this section the coding results of the proposed approach is reported.

The aspects verified in this part are the following:
\begin{itemize}
\item The upsampling filter separability and the symmetrisation impact on coding performance is measured
\item As the number of parameters for kernel filters is significantly reduced , the impact of extending the number of taps and the number of filters is considered
\item The coding gains are measured after the inclusion of the additional filtering branch
\end{itemize}

\subsection{Experimental setup}

The coding setup considers the encoding of the 41 sequences of the CLIC20 pro validation dataset~\cite{clic20pro}. The distortion metric is chosen as the mean squared error, and a rate constraint $\lambda={0.0001, 0.0004, 0.0010, 0.0040, 0.0200}$ is added to explore the bit per pixel range of approximatively 0.05 to 1.00 bit per pixel.

The results are reported against two references: the \cc version 3.2 (CC3.2) with a decoder having 298 MAC per decoded pixel and versus the HEVC standard under its HM16.20 reference implementation.

The decoding complexity is set to approximatively 300 MAC following~\cite{blard}. Here, the autoregressive module has a context of 8 pixels and the first synthesis hidden layer has 8 neurons.

The results are reported in BD-rate, that is the amount of bits saved (negative value) for a given solution w.r.t. an anchor solution.

\begin{table}
\begin{center}
    \includegraphics[width=\columnwidth]{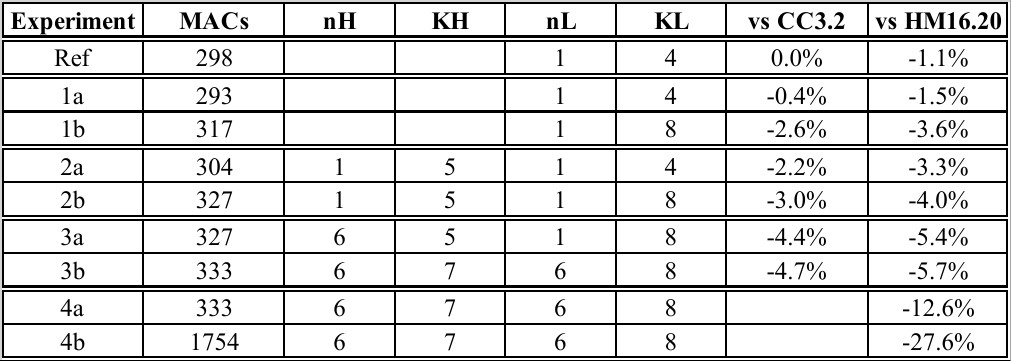}
	\caption{Experiment 1 measures the impact of the separability and filter symmetrisation. Experiment 2 quantifies the additional filter branch gains. Experiment 3 combines both contributions and extends the number of filters and kernels. Experiment 4 reports the results after full convergence for two decoding complexities.}
\label{table:TableBD}
\end{center}
\end{table}

The results are displayed in table~\ref{table:TableBD} in fast preset of~\cite{ccopen}  (10k training iterations) for  experiments 1 to 3. Experiment 4 extends the training to 100k iterations (slow mode) to achieve the best performance.
Parameters $n_H$, $K_H$, $n_L$ and $K_L$ stand for the number and kernel size for the $H$ and $L$ filters respectively.

\subsection{Experiment 1: impact of separability}

To measure the impact of separability, in a first experiment, a single upsampling kernel is used as in the legacy non-separable version.

The impact of separable and symmetrical filters is first evaluated with 4 and 8 taps. With $K=4$ taps one sees that the proposed approach provides a small improvement: indeed the additional constraints have no impact on the performance but, as the number of transmitted parameters is reduced from 16 to 2, a slight improvement is obtained. For $K=8$, the gains are noticeable, -2.6\% compared to the \cc reference. These gains were not possible with non-separable structures  as the transmitted number of parameters counterbalanced the effect of filter lengthening.
An example of frequency response is included in figure~\ref{fig:cc_freqresponse}e.

\subsection{Experiment 2 and 3: impact of the additional branch}

The second experiment measures the effect of adding the filter branch $H(z)$ on top of experiments 1a and 1b. It shows that one single filter ($n_H=1$) with $K_H=5$ taps brings gains for both cases.

In 3a and 3b the number of kernels is extended to 6. Also a slight benefit is noticed when the kernel size $K_H=7$. Gains are provided using both contributions, the additional filters in 3a provides approx. 1.4\% gain w.r.t. experiment 2b. Experiments 3b demonstrates that small additional benefits can  be seen using more filters and longer filters. Gains of around 4.5\% are demonstrated relative to the \cc 3.2 version.

\subsection{Experiment 4: overall performance}

Experiments 1-3 reported the performance of the improved \cc with 10k training iterations. Experiment 4a and 4b extends the learning process to 100k iterations for the 300 MAC range (4a) and the more complex operating point (4b), where the entropy neural network is extended to 16 neurons and the synthesis to 32. Compared to HEVC, respectively 12.6\% and 27.6\% of bit per pixel reduction is reported.

\begin{figure}
\begin{center}
    \includegraphics[width=\columnwidth]{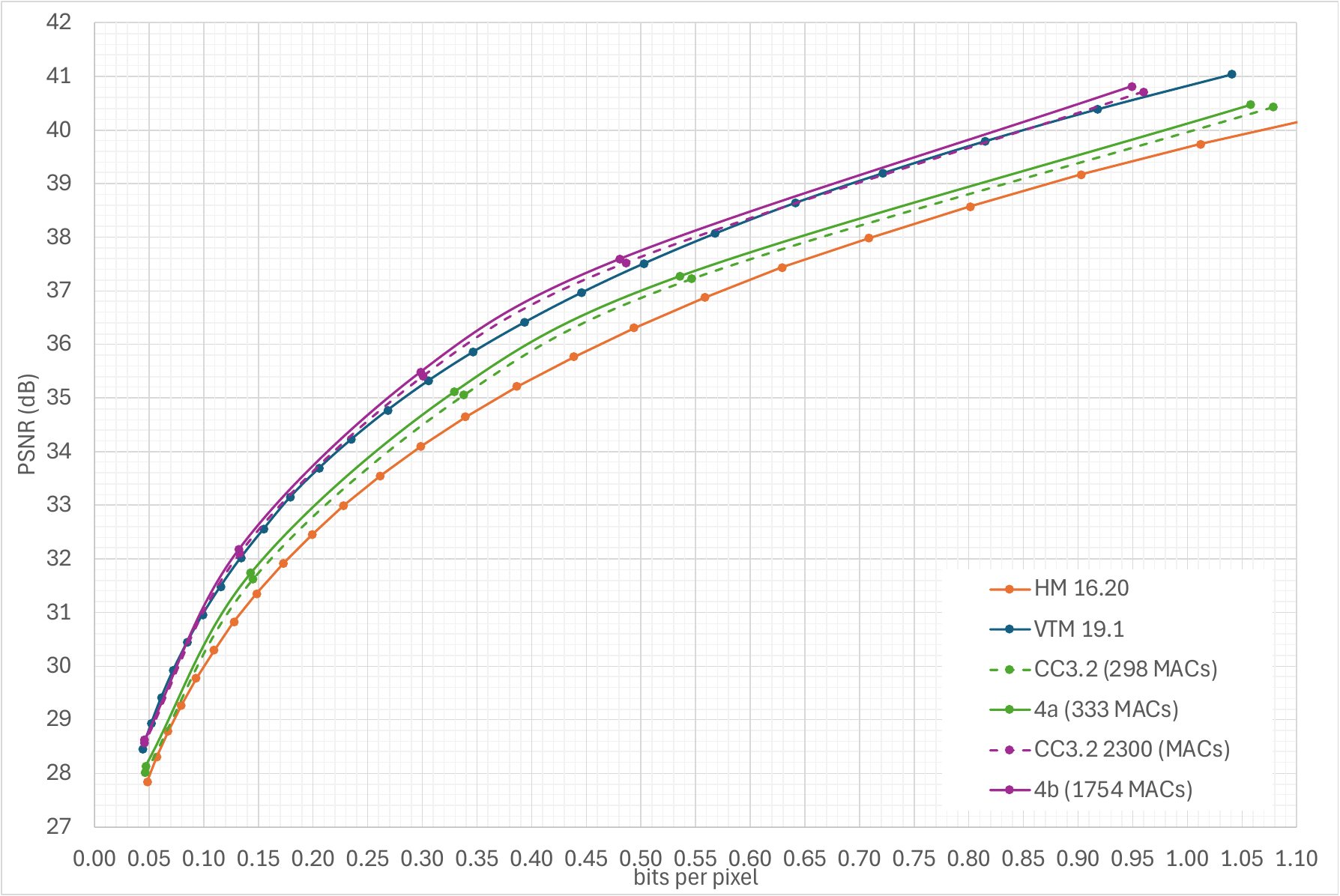}
	\caption{Overall rate distortion performance compared to previous \cc versions and the VTM}
\label{fig:vtm}
\end{center}
\end{figure}

Figure~\ref{fig:vtm} displays the rate distortion performance for the improved \cc version compared to its previous release (version 3.2), HEVC and VVC. The contributions on the upsampling process consistently improves the performance for both complexity points: -3.7\% and -2.6\% BD-rate gains achieved with the proposed modifications.

The modifications to the upsampling process allow \cc to strengthen its competitiveness relative to the VTM, as it progresses from -3.0\% to -5.5\% and more generally increases the attractiveness of overfitted image codecs.


\section{Conclusion}

Overfitted neural codecs need to adapt the neural representation to the signal in a compact fashion, while maintaining a low complexity. This paper demonstrates the effectiveness of using separable and symmetric filters in the upsampling process with no negative impact on the level of performance.

Moreover, the compactness of the filtering structure allows the extension of  both the filtering kernels and the number of filters to allow more flexibility. This results in a better adaptation of the neural representation to the images and leads to compression improvements of the \cc codec by 5\% at lower complexities and more than 2\% for the higher operating modes.

This work is made open source~\cite{ccopen}.

\bibliographystyle{IEEEbib}

\bibliography{refs}

\begin{thebibliography}{10}

\bibitem{ccopen}
{Orange Open Source},
\newblock ``Cool-chic open repository,''
\newblock 2024,
\newblock https://github.com/Orange-OpenSource/Cool-Chic.

\bibitem{hevc}
Gary~J. Sullivan, Jens-Rainer Ohm, Woo-Jin Han, and Thomas Wiegand,
\newblock ``Overview of the high efficiency video coding (hevc) standard,''
\newblock {\em IEEE Transactions on Circuits and Systems for Video Technology},
  vol. 22, no. 12, pp. 1649--1668, 2012.

\bibitem{10.5555/559856}
David~S. Taubman and Michael~W. Marcellin,
\newblock {\em JPEG 2000: Image Compression Fundamentals, Standards and
  Practice},
\newblock Kluwer Academic Publishers, USA, 2001.

\bibitem{10123093}
Jo\~ao Ascenso, Elena Alshina, and Touradj Ebrahimi,
\newblock ``The {JPEG AI} standard: Providing efficient human and machine
  visual data consumption,''
\newblock {\em IEEE MultiMedia}, vol. 30, no. 1, pp. 100--111, 2023.

\bibitem{balle}
Johannes Ball\'e, David Minnen, Saurabh Singh, Sung~Jin Hwang, and Nick
  Johnston,
\newblock ``Variational image compression with a scale hyperprior,''
\newblock in {\em International Conference on Learning Representations}, 2018.

\bibitem{dupont2021coincompressionimplicitneural}
Emilien Dupont, Adam Goliński, Milad Alizadeh, Yee~Whye Teh, and Arnaud
  Doucet,
\newblock ``{COIN}: Compression with implicit neural representations,'' 2021.

\bibitem{coolchic}
Th\'eo Ladune, Pierrick Philippe, F\'elix Henry, Gordon Clare, and Thomas
  Leguay,
\newblock ``Cool-chic: Coordinate-based low complexity hierarchical image
  codec,''
\newblock in {\em Proceedings of the IEEE/CVF International Conference on
  Computer Vision (ICCV)}, October 2023, pp. 13515--13522.

\bibitem{blard}
Th\'eophile Blard, Th\'eo Ladune, Pierrick Philippe, Gordon Clare, Xiaoran
  Jiang, and Olivier D\'eforges,
\newblock ``Overfitted image coding at reduced complexity,''
\newblock in {\em 2024 European Signal Processing Conference ({EUSIPCO})},
  2024.

\bibitem{9503377}
Benjamin Bross, Ye-Kui Wang, Yan Ye, Shan Liu, Jianle Chen, Gary~J. Sullivan,
  and Jens-Rainer Ohm,
\newblock ``Overview of the versatile video coding ({VVC}) standard and its
  applications,''
\newblock {\em IEEE Transactions on Circuits and Systems for Video Technology},
  vol. 31, no. 10, pp. 3736--3764, 2021.

\bibitem{c3}
Hyunjik Kim, Matthias Bauer, Lucas Theis, Jonathan~Richard Schwarz, and Emilien
  Dupont,
\newblock ``C3: High-performance and low-complexity neural compression from a
  single image or video,''
\newblock in {\em Proceedings of the IEEE/CVF Conference on Computer Vision and
  Pattern Recognition (CVPR)}, June 2024, pp. 9347--9358.

\bibitem{10533758}
Th\'eo Ladune, Pierrick Philippe, Gordon Clare, F\'elix Henry, and Thomas
  Leguay,
\newblock ``Cool-chic: Perceptually tuned low complexity overfitted image
  coder,''
\newblock in {\em 2024 Data Compression Conference (DCC)}, 2024, pp. 565--565.

\bibitem{clic20pro}
``Challenge on learned image coding 2020,''
\newblock 2020,
\newblock http://clic.compression.cc/2021/tasks/.

\end{thebibliography}

\end{document}